\documentclass[11pt]{article}

\usepackage[final]{acl}

\usepackage{times}
\usepackage{latexsym}

\usepackage[T1]{fontenc}

\usepackage[utf8]{inputenc}

\usepackage{microtype}

\usepackage{inconsolata}

\usepackage{graphicx}

\usepackage{tabularx}
\usepackage{makecell}
\usepackage{booktabs}
\usepackage{ragged2e}
\usepackage{multirow}
\usepackage{amsmath}

\newcolumntype{Y}{>{\RaggedRight\arraybackslash}X}

\title{Rethinking Fairness in LLM-Based Recommender Systems: A Survey}

\author{
Song-Duo Ma\thanks{Equal contribution.} \quad
Chu-Yun Chen\footnotemark[1] \quad
Bang-An Li\footnotemark[1] \\
\textbf{Pin-Yu Chen}\footnotemark[1] \quad
\textbf{Shau-Yung Hsu}\footnotemark[1] \quad
\textbf{Yun-Nung Chen} \\
National Taiwan University, Taipei, Taiwan \\
\texttt{\{r14944001, r14944023, r14922038\}@ntu.edu.tw} \\
\texttt{\{p14922004, r14922090\}@ntu.edu.tw, y.v.chen@ieee.org}
}

\begin{document}
\maketitle

\begin{abstract}
Large Language Models (LLMs) are reshaping recommender systems by enabling more semantic, generative, and interactive recommendation pipelines. However, this shift also introduces new fairness challenges, as biases may arise from pretrained knowledge, prompts, generated explanations, decoding strategies, and feedback loops. This survey provides a systematic review of fairness in LLM-based recommender systems (LLM4Rec), organizing existing studies through a two-dimensional view of bias mechanisms and fairness targets, together with a structured overview of the evaluation landscape and mitigation strategies. We further connect fairness with broader trustworthy concerns, including explainability, privacy, robustness, and controllability. To the best of our knowledge, this is the first survey specifically focused on fairness in LLM4Rec, aiming to provide a structured foundation for future research on comprehensive and reliable fairness evaluation in LLM4Rec.\footnote{\url{https://github.com/MiuLab/FairLMRec}}
\end{abstract}

\section{Introduction}
The emergence of Large Language Models (LLMs) is reshaping the design of recommender systems (RecSys), moving them beyond traditional collaborative filtering and user/item IDs toward recommendation pipelines enhanced by semantic understanding, natural language generation, and reasoning capabilities~\citep{lin2025how}. In LLM-based recommender systems (LLM4Rec), LLMs can be integrated as different modules, such as extracting user preferences and item features, re-ranking candidate items, and generating rationales~\citep{wang2024towards}. Therefore, LLMs help address long-standing limitations of traditional RecSys, including data sparsity, cold-start problems, and limited explainability.

However, the integration of LLMs also introduces new fairness challenges~\citep{gptfair2023, understandingbiases2025}. Fairness has long been an important issue in RecSys: user-side fairness focuses on equitable recommendation quality across user groups~\citep{uncertaintyfairness2026, cfairllm2025}, while item-side fairness focuses on whether different items or providers receive fair exposure~\citep{item-sidefair2024, gptnews2023}. In LLM4Rec, fairness becomes more complex because bias may arise not only from interaction data and exposure distributions, but also from prompt design, pretrained knowledge, recommendation explanations, and generation processes. Although recent studies have begun to examine these issues, the literature remains fragmented across different bias mechanisms, fairness targets, evaluation resources, and mitigation strategies, making it difficult to obtain a systematic understanding of fairness in LLM4Rec.

To address this gap, this paper aims to systematically review and discuss fairness issues in LLM4Rec. We first introduce the transition from traditional RecSys to LLM4Rec and summarize the major roles that LLMs play in the recommendation pipeline. We then classify fairness problems in LLM4Rec along two dimensions, bias mechanisms and fairness targets. Next, we review commonly used datasets in RecSys, existing fairness evaluation protocols, widely adopted fairness metrics, and the trade-off between utility and fairness~\citep{gptfair2023, item-sidefair2024}, followed by a discussion of fairness mitigation methods in LLM4Rec~\citep{fairnessbeprompted2026, fairaugment2025}. Finally, we examine the connections between fairness and other trustworthy issues, including explainability, privacy, robustness, and controllability~\citep{privacypreserving2025, llmasarecommende2026}, providing directions for future research and evaluation.

Overall, our contributions are threefold. (1) We provide the first focused and systematic survey of fairness in LLM4Rec, unifying fragmented findings across RecSys, NLP, and trustworthy AI. (2) We propose a two-dimensional taxonomy of fairness issues by bias mechanisms and fairness targets, revealing not only well-studied directions but also critical blind spots in the literature. (3) We summarize evaluation resources, protocols, mitigation strategies, and cross-cutting trustworthy issues, establishing a foundation for holistic fairness evaluation in future LLM4Rec research.

\section{Background}
In recent years, as LLMs have become increasingly powerful, growing research has begun to integrate them into RecSys. LLMs can serve as specific modules within the recommendation pipeline and can also directly generate recommendation results. However, with the integration of LLMs, fairness issues that have long been studied in traditional RecSys have also taken on new forms. Bias no longer arises only from user interaction data, exposure distributions, or historical behavioral patterns, but may also originate from pretrained knowledge, prompt design, and generation processes.

\subsection{From Traditional RecSys to LLM4Rec}
Traditional RecSys rely predominantly on collaborative filtering and user/item IDs to capture behavioral patterns. While effective, these ID-based paradigms often struggle with data sparsity, cold-start problems, and limited interpretability. The transition to LLM4Rec marks a paradigm shift from purely ID-driven matching to semantic-driven reasoning~\citep{lin2025how}. By leveraging massive pretrained world knowledge and natural language understanding, LLMs can incorporate item descriptions, user reviews, and conversational context into recommendation pipelines, thereby enriching user and item representations and supporting more flexible recommendation scenarios~\citep{wang2024towards}. However, this shift also changes the sources of recommendation bias, as outcomes may be shaped not only by interaction data but also by pretrained knowledge, prompts, and generated language.

\subsection{The Roles of LLMs in Recommendation}

LLMs are increasingly embedded across recommendation pipelines due to their strong language understanding, semantic representation, and reasoning abilities~\citep{wang2024towards}. Depending on their position in the pipeline, LLMs can support different stages of recommendation.

\paragraph{User and Item Extractors}
LLMs can extract user preferences, item attributes, and contextual signals from interaction histories and item content~\citep{RLMRec2024, llmrec2024}. In contrast to traditional RecSys, which mainly rely on user/item IDs or collaborative filtering signals, LLMs encode unstructured information into more interpretable semantic representations. This strengthens user/item representations and semantic understanding, particularly in cold-start scenarios.

\paragraph{Re-Rankers}
LLMs are used in the re-ranking stage after candidate items have been retrieved or generated~\citep{llmrank2024, ur4rec2025}. Rather than serving as recommendation generators, LLMs reassess candidate relevance based on user preferences, interaction histories, and item descriptions. This makes them useful refinement modules for improving candidate ordering and explanation generation within recommendation pipelines.

\paragraph{Generators}
LLMs directly generate recommendation lists and rationales~\citep{p5_2022, M6Rec2022}. In this setting, recommendation is transformed into a natural language generation task, where LLMs generate results based on context, user history, and task requirements. Thus, generators are especially suitable for scenarios that require both recommendation results and explanations.

\paragraph{Explanation Modules}
LLMs can serve as explanation modules that generate natural language rationales for recommendation decisions~\citep{xrec2024, llmexplanation2024}. They can explain before recommendation why certain items should be included in the candidate set, or explain after recommendation how the final results match user preferences. This capability improves the transparency, explainability, and trustworthiness of RecSys.

\begin{table*}[t]
\centering
\scriptsize
\setlength{\tabcolsep}{5pt}
\renewcommand{\arraystretch}{1.18}
\begin{tabularx}{\textwidth}{>{\RaggedRight\arraybackslash}p{3cm} Y Y Y}
\toprule
\textbf{Bias Mechanism} & \textbf{User-Side Fairness} & \textbf{Item-Side Fairness} & \textbf{Two-Sided Fairness} \\
\midrule

\textbf{Social and Attribute Bias}
&
\citet{gptfair2023}, \citet{cfairllm2025}, \citet{normative2024}, \citet{fairnessmatter2024}, \citet{up52024}, \citet{faireval2025}, \citet{facter2025}, \citet{improvefairness2025}, \citet{uncertaintyfairness2026}, \citet{comparative2025}, \citet{fairnessidentification2025}, \citet{lightweight2026}, \citet{multiattribute2025}
&
\emph{Rarely studied explicitly}
&
\citet{academicbias2025}, \citet{fairwork2025} \\
\midrule

\textbf{Linguistic and Knowledge Bias}
&
\citet{towardunderstandbias2023}, \citet{studyunfairness2024}, \citet{fairnessbeprompted2026}, \citet{revealbiases2025}
&
\citet{bifair2025}, \citet{llmrecg2025}, \citet{item-sidefair2024}
&
\citet{investigatingmitigatingunfairness2025} \\
\midrule

\textbf{Data and Propensity Bias}
&
\citet{mitigatingpropensity2025}, \citet{fairaugment2025}, \citet{towardfair2026}, \citet{towardunderstandbias2023}
&
\citet{item-sidefair2024}
&
\citet{unveilbias2026}, \citet{semanticandbias2026}, \citet{leadfairrec2025}, \citet{deconflating2026} \\
\midrule

\textbf{System and Optimization Bias}
&
\citet{llmfocus2026}, \citet{iagent2025}
&
\citet{understandingbiases2025}, \citet{gptnews2023}, \citet{sprec2025}, \citet{dualdebiasing2025}, \citet{split2026}, \citet{bifair2025}, \citet{llmasarecommende2026}, \citet{decoding2024}, \citet{collabrec2026}, \citet{refining2026}, \citet{dubash2026self}
&
\citet{polarization2026}, \citet{algorithmic2025}, \citet{echoes2026} \\
\bottomrule
\end{tabularx}
\caption{A two-dimensional taxonomy of fairness research in LLM4Rec, organized by fairness target and bias mechanism. The taxonomy reveals where current studies are concentrated and highlights underexplored intersections across bias sources and fairness objectives.}
\label{tab:fairness_taxonomy_2d}

\end{table*}

\subsection{Fairness Challenges in LLM4Rec}

Fairness has long been an important issue in RecSys because recommendations shape users' access to information, products, services, and opportunities. In LLM4Rec, these concerns become more complex because unfairness may arise not only from historical interactions and exposure imbalance, but also from pretrained knowledge, prompt design, and candidate ordering~\citep{understandingbiases2025}. Therefore, fairness in LLM4Rec should be examined as a system-level property across the pipeline, rather than only as a static ranking objective.

\section{A Taxonomy of Fairness in LLM4Rec}
\label{sec:taxonomy}

Fairness issues in LLM4Rec vary in both their sources and affected stakeholders. To understand these differences, we analyze fairness issues in LLM4Rec along two dimensions as shown in Table~\ref{tab:fairness_taxonomy_2d}: \textit{bias mechanisms} and \textit{fairness targets}. Bias mechanisms examine the sources from which unfairness emerges, which can be broadly categorized into four types. Fairness targets describe the stakeholders affected by recommendation outcomes, including users, items, and both sides jointly. 

\begin{figure*}[t]
    \centering
    \includegraphics[width=\textwidth]{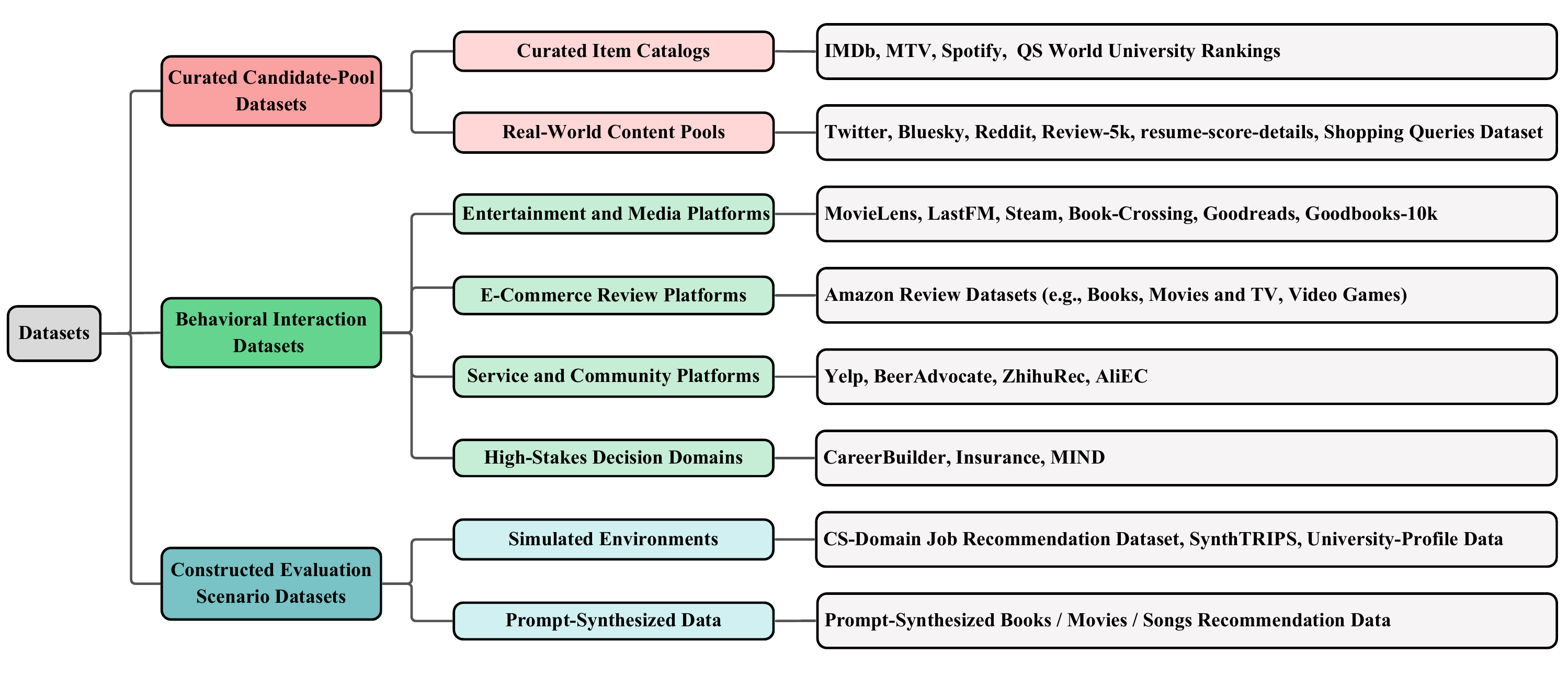}
    \caption{Taxonomy of datasets used in fairness-oriented LLM4Rec research. Datasets are organized into three categories: curated candidate-pool datasets, behavioral interaction datasets, constructed evaluation scenario datasets.}
    \label{fig:dataset}
\end{figure*}

\subsection{Bias Mechanisms}

Bias mechanisms describe where unfairness emerges in LLM4Rec, from data and exposure patterns to prompts, generation, and pipeline design.

\paragraph{Social and Attribute Bias}

Social and attribute bias arises when recommendations vary across sensitive or socially salient attributes, such as gender, age, nationality, religion, occupation, or race. In LLM4Rec, such bias may appear through explicitly provided attributes or through attributes inferred from names, occupations, language styles, or conversational contexts~\citep{gptfair2023, cfairllm2025}. While most studies focus on user-side demographic disparity, this mechanism can also affect two-sided fairness in institutional or labor-market matching scenarios~\citep{fairwork2025}.

\paragraph{Linguistic and Knowledge Bias}
Linguistic and knowledge bias originates from language patterns, cultural associations, and world knowledge encoded in LLM pretraining corpora. Even without explicit demographic signals, LLM4Rec may associate names, locales, professions, or writing styles with particular preferences, groups, or cultural contexts~\citep{towardunderstandbias2023, studyunfairness2024}. Such semantic priors may further bias recommendations toward mainstream or culturally dominant items over niche, local, or underrepresented alternatives.

\paragraph{Data and Propensity Bias}
Data and propensity bias stems from popularity skews, selection effects, exposure inequalities, and imbalanced interaction histories. LLM4Rec may inherit these biases when LLMs are used as rankers, user modelers, item encoders, or generators~\citep{mitigatingpropensity2025, fairaugment2025}. In generative settings, this can appear as repeated recommendation of popular or historically exposed items, suppressing long-tail alternatives~\citep{item-sidefair2024}.

\paragraph{System and Optimization Bias}
System and optimization bias arises when design and inference choices systematically shape recommendation outcomes. In LLM4Rec, fairness can be affected not only by model knowledge or training data, but also by how prompts are formulated, candidates are ordered, outputs are decoded, and user feedback is incorporated over time~\citep{understandingbiases2025, decoding2024, echoes2026}. This mechanism is important because it spans multiple stages of the pipeline, making system design not only a source of unfairness but also a practical point of intervention.
 
\subsection{Fairness Targets}
Fairness in recommendation is fundamentally a multi-stakeholder objective. In our taxonomy, we classify fairness goals based on the specific stakeholders affected by the recommendation outcomes.

\paragraph{User-Side Fairness} 
This dimension ensures equitable recommendation quality and utility across different user groups. It primarily aims to mitigate demographic bias regarding sensitive attributes such as gender, age, and race \citep{gptfair2023}. User-side fairness is typically evaluated from two perspectives: group fairness, which examines whether recommendation performance is comparable across demographic groups, and individual fairness, which requires similar users to receive similar treatment~\citep{cfairllm2025}.

\paragraph{Item-Side Fairness} 
This objective centers on the equitable allocation of visibility among items or content providers. It specifically addresses popularity bias and exposure disparity, ensuring that long-tail items, niche content, and underrepresented providers are not unfairly marginalized by the system's ranking distribution \citep{item-sidefair2024}.

\paragraph{Two-Sided Fairness} 
Two-sided fairness considers the interests of both users and items or providers, aiming to balance user-side recommendation utility with item-side exposure equity~\citep{leadfairrec2025}. In LLM4Rec, this objective is particularly challenging because improving personalization for users may unintentionally amplify exposure disparities among items or providers.

\subsection{Taxonomy Assignment and Coverage}
\label{subsec:assignment}

\paragraph{Assignment Rules}
Each study is assigned to a single primary cell in Table~\ref{tab:fairness_taxonomy_2d}, based on the bias source it mainly evaluates or mitigates and the stakeholder outcome it directly measures. User-side fairness covers a disparity in user utility or treatment; item-side fairness covers item or provider exposure and opportunity; and two-sided fairness requires both sides to be jointly evaluated or jointly optimized, not merely discussed. \emph{Rarely studied explicitly} denotes a lack of studies whose primary empirical focus falls in that intersection, rather than a complete absence of related work.

\paragraph{Interpreting Sparse Cells}
The uneven distribution in Table~\ref{tab:fairness_taxonomy_2d} also reflects what can currently be measured. Item-side social bias is hard to study explicitly because provider-side protected attributes are rarely available and provider benefit is domain-dependent. Two-sided linguistic and knowledge bias additionally requires jointly measuring user utility and provider benefit while separating pretrained semantic priors from popularity effects. We therefore interpret sparse cells as measurement constraints rather than as indicators of importance.

\section{Evaluation Resources and Protocols}
RecSys rely heavily on large amounts of data to model user preferences and item characteristics, making dataset properties central to fairness evaluation. We organize commonly used datasets into three main types based on their data and scenarios. Moreover, as discussed in Section~\ref{sec:taxonomy}, fairness issues in LLM4Rec can arise from various sources, making a single evaluation setting insufficient for comprehensively measuring RecSys fairness. To capture different forms of fairness issues, existing studies have adopted diverse evaluation protocols and metrics. We categorize these approaches into four groups based on their main evaluation focus.

\begin{figure*}[t]
    \centering
    \includegraphics[width=\textwidth]{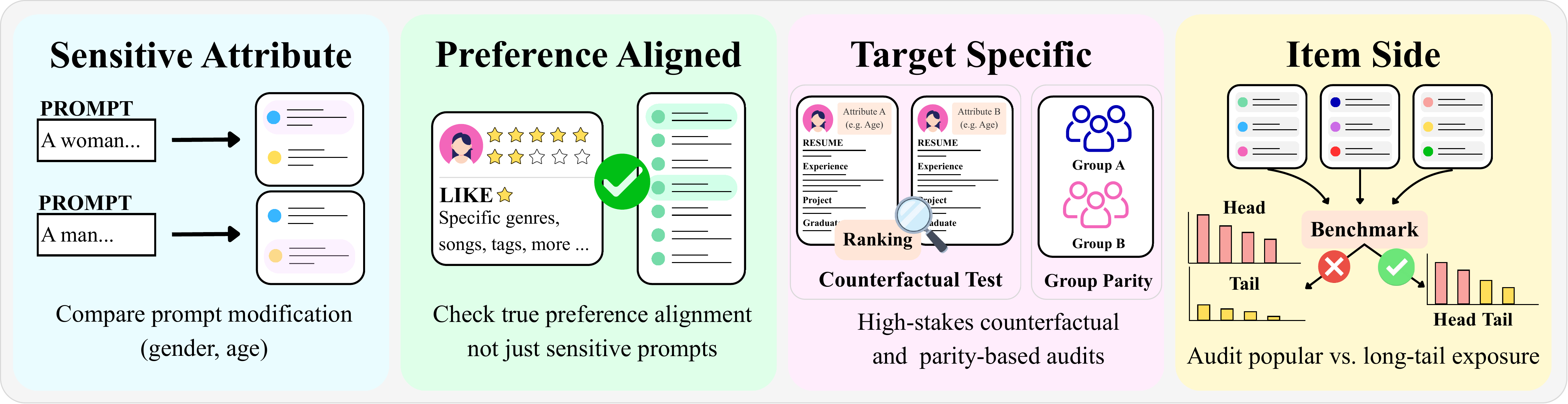}
    \caption{Taxonomy of fairness evaluation protocols in LLM4Rec. The protocols are categorized into four families based on their evaluation targets: Sensitive Attribute, Preference Aligned, Target Specific, and Item Side.}
    \label{fig:benchmark}
\end{figure*}

\subsection{Datasets and Data Sources}
\label{sec:datasets}

Empirical fairness research in LLM4Rec draws on diverse data sources. As summarized in Figure~\ref{fig:dataset}, we categorize existing datasets into three broad types according to how the evaluation data are constructed and used. Appendix~\ref{sec:appendix_dataset} further provides a dataset-level mapping between individual datasets and the corresponding studies that use them.

\paragraph{Curated Candidate-Pool Datasets}
Curated candidate-pool datasets are used to evaluate LLM recommendation behavior under controlled candidate sets. They include curated item catalogs, such as IMDb, MTV, and Spotify~\citep{faireval2025, uncertaintyfairness2026, revealbiases2025}, as well as real-world content pools collected from social media, reviews, and shopping queries~\citep{polarization2026, llmasarecommende2026}. By fixing the candidate pool while varying user profiles, prompts, or candidate descriptions, these datasets enable controlled analysis of whether LLMs exhibit biased ranking preferences across demographic, cultural, institutional, or task-specific contexts~\citep{gptfair2023, academicbias2025}.

\paragraph{Behavioral Interaction Datasets}
Behavioral interaction datasets consist of observed user--item feedback, such as ratings, reviews, clicks, and purchase histories, and serve as the empirical foundation for many LLM4Rec fairness studies. These datasets span entertainment and media platforms~\citep{understandingbiases2025, item-sidefair2024, fairnessmatter2024}, e-commerce review platforms~\citep{bifair2025, fairaugment2025}, service and community platforms, and high-stakes decision domains such as employment and insurance recommendation~\citep{fairwork2025, studyunfairness2024}. They are commonly used to evaluate user-side disparities, item-side exposure inequality, popularity bias, and the trade-off between recommendation utility and fairness.

\paragraph{Constructed Evaluation Scenario Datasets}
Constructed evaluation scenario datasets are designed to evaluate fairness under controlled recommendation settings rather than to model naturally occurring interaction logs. They include simulated environments for domains such as job, tourism, and academic recommendation~\citep{deconflating2026, collabrec2026, academicbias2025}, as well as prompt-synthesized recommendation data for books, movies, and songs~\citep{unveilbias2026}. By explicitly specifying user profiles, task contexts, or stakeholder roles, these datasets are especially useful for studying LLM-specific fairness issues, including prompt sensitivity, counterfactual user profiles, and scenario-level bias.

\subsection{Fairness Evaluation in LLM4Rec}
\label{sec:evaluation}

As illustrated in Figure~\ref{fig:benchmark}, existing fairness evaluation protocols for LLM4Rec can be organized into four categories according to their primary evaluation focus: sensitive attributes, preference-aligned benefit, target-specific decisions, and item-side exposure. These categories differ in their evaluation settings, assumptions, and adopted fairness metrics. A detailed mapping of metric families and representative studies, together with a comparative analysis of the assumptions, strengths, limitations, and suitable settings of these protocols, is provided in Appendix~\ref{sec:appendix_protocols}.

\paragraph{Sensitive Attribute}
Sensitive-attribute evaluation protocols evaluate fairness by modifying sensitive attributes, such as gender or age, in prompts and measuring whether the recommendation outputs change. FaiRLLM~\citep{gptfair2023} and its extensions~\citep{faireval2025,uncertaintyfairness2026,revealbiases2025} follow this perturbation-based paradigm. The corresponding metrics are usually similarity- or ranking-based, including SNSR and SNSV~\citep{gptfair2023}, as well as Jaccard@K, SERP$^{*}$@K, and PRAG$^{*}$@K. These metrics capture how much the recommendation list changes under sensitive-attribute perturbations. However, output differences do not necessarily imply unfairness, as they may reflect legitimate preference differences rather than biased treatment.

\paragraph{Preference Aligned}
Preference-aligned evaluation protocols evaluate whether output differences actually harm user benefit, rather than simply measuring list similarity. CFaiRLLM~\citep{cfairllm2025} and the normative framework of \citet{normative2024} follow this direction by assessing whether users with different sensitive attributes receive comparable recommendation benefits. The main metric is benefit deviation, such as $\Delta B$~\citep{cfairllm2025,normative2024}, which measures whether recommendation changes lead to utility loss for specific user groups. Compared with sensitive-attribute protocols, this approach better reflects user welfare, but it depends on accurate preference and benefit estimation.

\paragraph{Target Specific}
Target-specific evaluation protocols focus on domain-specific recommendation scenarios, such as job and academic recommendation. Fairness is typically evaluated through counterfactual testing, group-level parity, and domain-specific ranking outcomes. Prior studies show that bias may arise from demographic attributes or proxy cues~\citep{studyunfairness2024}, while frameworks such as FairWork~\citep{fairwork2025} and JobRec~\citep{deconflating2026} provide more realistic evaluation settings. The corresponding metrics adapt classification-based fairness measures to ranking scenarios, including Statistical Parity (SP), Equal Opportunity (EO), and PPV\_diff~\citep{fairwork2025}. Additional metrics such as DRS, GRS, and U-NDCG capture domain-specific disparities in representation and ranking fairness~\citep{academicbias2025,studyunfairness2024}. These evaluation protocols are rigorous but less generalizable across domains.

\paragraph{Item Side}
Item-side evaluation protocols examine whether exposure is equitably distributed across items, especially between popular and long-tail items. Prior studies measure popularity bias, provider-side disparities, and long-tail suppression in LLM4Rec~\citep{understandingbiases2025,gptnews2023,item-sidefair2024}. The corresponding metrics include inequality-based measures such as the Gini Index, Herfindahl-Hirschman Index (HHI), and entropy~\citep{understandingbiases2025}, group-level exposure metrics such as MGU and DGU~\citep{item-sidefair2024}, and long-tail coverage metrics~\citep{semanticandbias2026}. However, improving item-side exposure fairness may conflict with recommendation relevance.

\paragraph{Trade-Offs}
Fairness improvements often trade off with utility metrics such as NDCG and hit ratio. For example, reducing popularity bias may increase long-tail exposure but lower relevance, while enforcing group-level parity may weaken personalization. Although prompting, reasoning, and debiasing strategies can sometimes mitigate these trade-offs~\citep{deconflating2026}, evaluation protocols should report both fairness and utility metrics to reflect practical LLM4Rec performance.

\subsection{Evaluation Gaps and Limitations}

Despite recent progress, fairness evaluation in LLM4Rec remains limited by inconsistent fairness definitions, heavy reliance on synthetic prompting, and the lack of realistic interactive evaluation settings. Existing protocols often capture only isolated aspects of fairness, making cross-study comparison difficult because results depend strongly on model choice, prompt templates, candidate construction, decoding settings, and sensitive-group definitions. Moreover, many studies report only single-run results, despite the stochasticity of generative recommendation, leaving the stability of observed fairness disparities unclear. Current metrics are also largely adapted from traditional RecSys and may not fully capture open-ended generation, explanation faithfulness, or multi-turn effects. Finally, evaluations increasingly relying on LLM-based judgments may introduce additional evaluator bias unless validated against human or task-grounded criteria. These limitations highlight the need for standardized reporting, repeated-run analysis, and evaluation protocols tailored specifically to generative and interactive LLM4Rec.

\section{Fairness Mitigation in LLM4Rec}
\label{sec:mitigation}

Fairness mitigation in LLM4Rec requires targeted interventions across the recommendation pipeline, addressing biases that arise from data, prompts, model behavior, and ranking decisions. Building on the bias sources and impact scopes characterized in Section~\ref{sec:taxonomy}, these interventions can be understood as operating at different stages of the pipeline. We organize them into four levels: input, data, model, and output mitigation. A comparative analysis of these intervention levels, including their assumptions, strengths, limitations, and suitable deployment settings, is provided in Appendix~\ref{sec:appendix_mitigation_comparison}.

\subsection{Input-Level Mitigation}
Input-level mitigation guides LLMs toward fair outcomes during inference without altering parameters. Approaches include online prompt optimization techniques that iteratively balance diversity and popularity bias \citep{split2026}. Additionally, conformal thresholding combined with prompt engineering can progressively repair adversarial prompts and guarantee fairness criteria \citep{facter2025}. However, prompting can be brittle, as its effectiveness depends on model reasoning ability and domain complexity~\citep{fairnessbeprompted2026}.

\subsection{Data-Level Mitigation}
Data-level mitigation targets historical biases in interaction logs, such as propensity and popularity skews. Counterfactual data augmentation perturbs user histories to balance the representation of diverse item groups \citep{fairaugment2025}. For high-stakes domains, counterfactual debiasing reduces the influence of popularity-driven interactions, improving both user utility and item exposure fairness~\citep{leadfairrec2025}. Moreover, causal intervention methods further model and mitigate inherited propensity bias, preventing LLMs from amplifying historical interaction patterns \citep{mitigatingpropensity2025}.

\subsection{Model-Level Mitigation}
Model-level mitigation improves fairness by updating or regularizing model parameters, while avoiding costly full-parameter retraining~\citep{towardfair2026}. Instead, Parameter-Efficient Fine-Tuning (PEFT) techniques, such as lightweight gated adapters and kernelized projections, dynamically control bias while preserving utility \citep{lightweight2026}. At the architecture level, foundation models can be explicitly regularized for fairness \citep{up52024}, and bi-level optimization can decouple utility objectives from fairness constraints \citep{bifair2025}. For intersectional biases, Mixture-of-Experts (MoE) combined with contrastive learning effectively assigns specialized modules to handle distinct sensitive attributes \citep{multiattribute2025}.

\subsection{Output-Level Mitigation}

Output-level mitigation adjusts recommendation outcomes during decoding or after generation without modifying the underlying model parameters. Autoregressive decoding can inadvertently amplify biases, such as worsening popularity skew due to length normalization artifacts. Modifying this process with debiasing-diversifying decoding (D$^3$) directly curtails representation homogeneity \citep{decoding2024}. Beyond text generation, post-hoc re-ranking acts as a secondary fairness filter. Recent works utilize LLMs as explainable re-rankers to audit and explicitly adjust candidate rankings \citep{explaiablereranker2025}, or apply dual debiasing frameworks via Inverse Propensity Score (IPS) to correct exposure disparities in generated lists \citep{dualdebiasing2025}.

\subsection{Key Takeaways Across Mitigation Strategies}

Mitigation strategies differ primarily in intervention depth and required system access. Input- and output-level methods are easier to deploy and remain applicable to closed-source or fixed-backbone systems, whereas data- and model-level methods can address deeper sources of bias but require greater access to training data, model parameters, or optimization procedures. These differences make the appropriate mitigation strategy highly dependent on the source of unfairness and deployment constraints.

Fairness and utility should also not be viewed as an inevitable trade-off. While exposure-oriented interventions may reduce relevance in some settings, correcting popularity bias, selection bias, or spurious correlations can improve fairness and recommendation utility simultaneously. Mitigation should therefore be evaluated jointly with fairness and utility metrics and interpreted relative to the targeted bias mechanism.

Finally, most existing methods are evaluated on static recommendation outputs. Whether fairness improvements persist across conversational feedback, repeated exposure, and multi-turn recommendation trajectories remains largely unexplored, making long-term and interactive fairness an important direction for future work.

\section{Cross-Cutting Trustworthy Issues}

Fairness in LLM4Rec is increasingly intertwined with other dimensions of trustworthiness. Explanations can reveal unfairness, but may also obscure or legitimize biased outcomes when they are misleading or unfaithful. Privacy constraints can alter how personalization quality is distributed across user groups, while robustness and controllability determine whether fairness persists under perturbations and can be reliably enforced across the recommendation pipeline. Therefore, fairness in LLM4Rec should be studied as a cross-cutting system property rather than as a standalone metric.

\subsection{Fairness and Explainability}

Explainability provides an important lens for fairness in LLM4Rec because natural language rationales can make biased recommendation behavior more visible. Unlike traditional RecSys, LLM4Rec can directly justify ranked lists, summarize user preferences, and support post-hoc auditing. Thus, explanations can serve as user interfaces and diagnostic signals of unfair recommendation patterns.

Existing studies mainly use explanations for three purposes. First, explanations can reveal whether recommendations are overly driven by popularity or historical interaction frequency, thereby supporting popularity debiasing~\citep{leverage2026}. Second, LLMs can act as explainable re-rankers, using textual justifications, candidate shuffling, or self-consistency to make ranking decisions more inspectable and reduce popularity or positional bias~\citep{explaiablereranker2025}. Third, explanations can be combined with counterfactual or causal debiasing, for example by filtering out-of-character history items or integrating causal debiasing with multimodal explanation generation~\citep{mitigatepopularitybias2025, LLM4Rec2025}.

However, explainability alone does not guarantee fairness. LLM-generated explanations may sound plausible while obscuring biased ranking factors such as popularity, stereotypes, or spurious correlations. Therefore, fairness-aware explainability should evaluate not only whether explanations are understandable, but also whether they are bias-aware, factually grounded, and faithful to the underlying recommendation process~\citep{mitigatingmisleadingness2025}.

\subsection{Fairness and Privacy}

Privacy introduces a distinct fairness concern in LLM4Rec because privacy protection can change how personalization quality is distributed across users and items. LLM4Rec often rely on rich user histories, profiles, and conversational contexts, which may expose sensitive preferences or enable private attribute inference. Even when explicit demographic attributes are removed, LLMs may still infer sensitive information from user text, such as gender, occupation, or other personal attributes~\citep{stereotypes2025, beyond2024}.

Existing studies suggest that privacy-preserving techniques, including differential privacy, federated learning, and pseudonymization, can reduce information leakage but may also weaken personalization signals~\citep{privacypreserving2025}. Such degradation may disproportionately affect underrepresented users, niche-preference groups, or long-tail items whose signals are already sparse. Moreover, the effects of privacy mechanisms on utility and bias can vary across models, datasets, privacy budgets, user groups, and item popularity groups~\citep{privacyutilitybias2025}. Therefore, fairness-aware privacy in LLM4Rec should move beyond the conventional trade-off between privacy and utility and evaluate whether privacy protection unevenly reshapes user-side utility, item-side exposure, and popularity bias.

\subsection{Fairness and Robustness}

Robustness is essential to fairness in LLM4Rec because fair behavior should not depend on a specific prompt, user description, context, or interaction trajectory. Since LLM4Rec are highly sensitive to natural language inputs and conversational signals, minor changes in wording, profiles, or contextual framing may lead to unstable fairness outcomes.

Existing studies show that fairness can vary across personas and demographic profiles~\citep{faireval2025}, emerge from implicit demographic cues even without explicit sensitive attributes~\citep{studyunfairness2024}, and be distorted by irrelevant contextual signals~\citep{llmfocus2026}. In interactive settings, feedback loops may further amplify earlier biases~\citep{echoes2026}, while recommender agents may be vulnerable to injected bias or preference manipulation~\citep{llmasarecommende2026}. Thus, robustness-aware fairness reframes fairness evaluation as a stability question: whether LLM4Rec remain fair under prompt perturbations, context shifts, feedback loops, and adversarial manipulation.

\subsection{Fairness and Controllability}

Controllability is closely related to fairness in LLM4Rec because fair behavior can be actively steered through prompts, model design, decoding strategies, and agentic policies. Unlike traditional RecSys, LLM4Rec expose multiple control points across pipelines, making fairness not only an evaluative property but also a controllable behavior.

Existing studies show that prompt design and system roles can affect provider fairness, temporal stability, recency bias, and high-stakes recommendation outcomes~\citep{understandingbiases2025, fairnessbeprompted2026}. Fairness can also be controlled through model-level objectives~\citep{up52024}, decoding strategies that reduce popularity bias or homogeneity~\citep{decoding2024}, and agentic mechanisms that act as fairness-aware shields between users and RecSys~\citep{iagent2025}. Therefore, fairness-oriented controllability concerns where and how fairness constraints can be imposed across the pipeline.

\section{Open Challenges and Future Directions}
\label{sec:future}

Building on the preceding sections, we identify four priorities for fairness-oriented LLM4Rec, emphasizing stronger evidence, shared benchmarks, tailored protocols, and consistent reporting.

\paragraph{Toward Cross-Target Fairness Analysis}
Research remains unevenly distributed across fairness targets and bias mechanisms. As shown in Table~\ref{tab:fairness_taxonomy_2d}, studies concentrate on user-side social and attribute bias, whereas item-side social bias and two-sided linguistic or knowledge bias remain underexplored. These gaps partly reflect the difficulty of defining provider attributes and socially meaningful item groups, and of measuring exposure or representation in natural-language outputs. Future work should also examine interactions across stakeholders: optimizing user utility may amplify provider exposure disparities, while reducing popularity bias may affect demographic groups unevenly~\citep{item-sidefair2024, leadfairrec2025}. Protocols should therefore jointly report groupwise utility, item and provider exposure, representational harms, and their trade-offs.

\paragraph{Toward LLM-Specific Fairness Benchmarks}
Many protocols remain static and single-turn, although LLM4Rec introduces prompt sensitivity, open-ended outputs, generated rationales, multi-turn feedback, and agentic actions~\citep{fairnessbeprompted2026, echoes2026}. Benchmarks should cover ranking, generative and explanation-based, conversational, and agentic settings, while documenting models, tasks, candidate pools, prompts, group definitions, languages, and utility ground truth. Metrics should be scenario specific: ranking settings require groupwise utility and exposure metrics; generative settings require language-bias, faithfulness, and prompt-sensitivity diagnostics; and conversational or agentic settings require trajectory-level fairness, retrieval or tool-use disparities, and feedback-loop stability.

\paragraph{Toward Holistic Trustworthy Evaluation}
Fairness should be treated as part of the broader trustworthy recommendation pipeline. Fluent but unfaithful explanations may legitimize biased outcomes or create an appearance of fairness without reflecting the ranking process~\citep{mitigatingmisleadingness2025}. System-level protocols should jointly evaluate outcomes, explanations, privacy--utility--bias effects~\citep{privacyutilitybias2025}, robustness, and controllability. To calibrate evidence strength, syntheses should distinguish peer-reviewed findings from preprints, while empirical reports should include utility baselines, fairness metrics, robustness checks, repeated-run variability, and utility--fairness trade-offs.

\paragraph{Toward Multilingual and Cultural Fairness}
The reviewed evidence is predominantly English-centric. A recent study suggests that fairness disparities can persist under multilingual prompt perturbations~\citep{uncertaintyfairness2026}. Generalizing beyond English requires more than translation, as languages and cultures differ in expressing preferences, identities, sensitive attributes, and fairness expectations. Future benchmarks should include diverse languages, culturally grounded fairness definitions, validated translations, and cross-lingual robustness tests for recommendation outputs.

\section{Conclusion}

This survey provides a systematic overview of fairness in LLM4Rec, covering bias mechanisms, fairness targets, evaluation resources, mitigation methods, and cross-cutting trustworthy issues. We show that fairness risks arise from both traditional recommendation biases and LLM-specific factors, including pretrained knowledge, prompts, and generation processes. Our taxonomy highlights gaps in item-side fairness, two-sided trade-offs, interactive settings, and holistic trustworthy evaluation. We hope this survey can serve as a foundation for future research on more comprehensive fairness evaluation and mitigation in LLM4Rec.



\section*{Limitations}

Given the broad scope of LLM4Rec, this survey focuses on studies in which fairness is explicitly examined as an evaluation objective, observed bias phenomenon, or mitigation target. As a result, works that use LLMs for recommendation but do not directly analyze fairness-related outcomes are discussed only when they provide necessary background or contextual support.

In addition, fairness in LLM4Rec is studied under diverse settings, including different recommendation tasks, datasets, prompting protocols, model access levels, and evaluation metrics. As a result, direct comparison across papers is inherently difficult and may be misleading without careful consideration of their task settings, evaluation protocols, and metric definitions. Instead of quantitatively aggregating results across heterogeneous studies, this survey emphasizes conceptual organization, recurring patterns, and open research gaps. We believe this perspective better reflects the current state of the field and can serve as a foundation for more standardized future evaluation.

\section*{Acknowledgments}
This work was financially supported by the National Science and Technology Council (NSTC) and the Featured Area Research Center Program within the framework of the Higher Education Sprout Project by the Ministry of Education in Taiwan, under Grants 112-2223-E-002-012-MY5, 115-2628-E-002-023-MY4, and 115L900901.

\bibliography{custom}

\appendix

\begin{table*}[t]
\centering
\small
\renewcommand{\arraystretch}{1.15}
\setlength{\tabcolsep}{5pt}

\begin{tabularx}{\textwidth}{
    >{\raggedright\arraybackslash}X
    >{\centering\arraybackslash}p{0.075\textwidth}
    >{\centering\arraybackslash}p{0.105\textwidth}
    >{\centering\arraybackslash}p{0.115\textwidth}
    >{\centering\arraybackslash}p{0.115\textwidth}
    >{\centering\arraybackslash}p{0.18\textwidth}
}
\toprule
\textbf{Survey}
& \textbf{Year}
& \textbf{LLM}
& \textbf{RecSys}
& \textbf{Fairness}
& \textbf{Cross-Cutting Trustworthiness} \\
\midrule

\citeauthor{comprehensivesurvey2022}
& 2022 & -- & $\bullet$ & $\circ$ & $\bullet$ \\

\citeauthor{surveyfairness2023}
& 2023 & -- & $\bullet$ & $\bullet$ & -- \\

\citeauthor{fairnessinrecommendation2023}
& 2023 & -- & $\bullet$ & $\bullet$ & -- \\

\citeauthor{surveyfairnessaware2023}
& 2023 & -- & $\bullet$ & $\bullet$ & $\circ$ \\

\citeauthor{fairnessinrecommendersystem2023}
& 2023 & -- & $\bullet$ & $\bullet$ & -- \\

\addlinespace[2pt]

\citeauthor{surveyfairnessllm2024}
& 2024 & $\bullet$ & -- & $\bullet$ & -- \\

\citeauthor{recommendersystemeraofllm2024}
& 2024 & $\bullet$ & $\bullet$ & $\circ$ & $\circ$ \\

\citeauthor{surveyllmrecommendation2024}
& 2024 & $\bullet$ & $\bullet$ & $\circ$ & -- \\

\citeauthor{biasfairness2024}
& 2024 & $\bullet$ & -- & $\bullet$ & -- \\

\citeauthor{trustworthyrecsys2024}
& 2024 & -- & $\bullet$ & $\circ$ & $\bullet$ \\

\citeauthor{surveytrusthworthyrecsys2024}
& 2024 & -- & $\bullet$ & $\circ$ & $\bullet$ \\

\citeauthor{wang2024towards}
& 2024 & $\bullet$ & $\bullet$ & $\circ$ & $\circ$ \\

\addlinespace[2pt]

\citeauthor{lin2025how}
& 2025 & $\bullet$ & $\bullet$ & $\circ$ & $\circ$ \\

\citeauthor{llm4recsurveyintergration2025}
& 2025 & $\bullet$ & $\bullet$ & $\circ$ & $\circ$ \\

\citeauthor{comprehensivesurveyllmpowered2025}
& 2025 & $\bullet$ & $\bullet$ & $\circ$ & $\circ$ \\

\midrule

\textbf{Ours}
& \textbf{2026}
& $\boldsymbol{\bullet}$
& $\boldsymbol{\bullet}$
& $\boldsymbol{\bullet}$
& $\boldsymbol{\bullet}$ \\

\bottomrule
\end{tabularx}

\caption{
Comparison with representative surveys in adjacent research areas.
A filled circle ($\bullet$) indicates a primary focus, an open circle
($\circ$) indicates meaningful but secondary coverage, and a dash (--)
indicates little or no substantive coverage. A dimension is considered
a primary focus when it constitutes a central stated objective and
receives substantial dedicated coverage.
}
\label{tab:survey_comparison}
\end{table*}

\begin{table*}[t]
\centering
\small
\begin{tabularx}{\textwidth}{p{0.12\textwidth} p{0.13\textwidth} p{0.25\textwidth} X}
\toprule
\textbf{Dataset Type} & \textbf{Subcategory} & \textbf{Dataset / Data Source} & \textbf{Studies Using This Dataset or Data Source} \\
\midrule

\multirow{6}{=}{\textbf{Curated Candidate-Pool Datasets}}
& \multirow{4}{=}{Curated Item Catalogs}
& IMDb
& \citep{gptfair2023, faireval2025, uncertaintyfairness2026, revealbiases2025} \\

& & MTV
& \citep{gptfair2023, faireval2025, uncertaintyfairness2026} \\

& & Spotify
& \citep{revealbiases2025} \\

& & QS World University Rankings
& \citep{revealbiases2025, academicbias2025} \\

\cmidrule(lr){2-4}

& \multirow{2}{=}{Real-World Content Pools}
& Twitter, Bluesky, Reddit posts
& \citep{polarization2026} \\

& & Review-5k, Shopping Queries Dataset, resume-score-details
& \citep{llmasarecommende2026} \\

\midrule

\multirow{29}{=}{\textbf{Behavioral Interaction Datasets}}
& \multirow{14}{=}{Entertainment and Media Platforms}
& MovieLens-100K
& \citep{investigatingmitigatingunfairness2025, comparative2025} \\

& & MovieLens-1M
& \citep{cfairllm2025, understandingbiases2025, item-sidefair2024, up52024, facter2025, fairaugment2025, improvefairness2025, towardfair2026, split2026, echoes2026, fairnessidentification2025, lightweight2026} \\

& & MovieLens-10M
& \citep{sprec2025} \\

& & MovieLens-20M
& \citep{semanticandbias2026} \\

& & MovieLens-25M
& \citep{fairnessmatter2024} \\

& & MovieLens Latest Small
& \citep{understandingbiases2025} \\

& & MovieLens, unspecified version
& \citep{normative2024} \\

& & LastFM-1K
& \citep{understandingbiases2025, cfairllm2025} \\

& & LastFM-360K
& \citep{fairaugment2025, improvefairness2025, fairnessidentification2025} \\

& & LastFM, unspecified version
& \citep{split2026, comparative2025} \\

& & Steam
& \citep{item-sidefair2024, sprec2025, llmrecg2025, mitigatingpropensity2025, refining2026} \\

& & Goodreads
& \citep{sprec2025, iagent2025} \\

& & BookCrossing
& \citep{investigatingmitigatingunfairness2025, leadfairrec2025} \\

& & Goodbooks-10k
& \citep{semanticandbias2026} \\

\cmidrule(lr){2-4}

& \multirow{9}{=}{E-Commerce Review Platforms}
& Amazon Books
& \citep{bifair2025, echoes2026, llmfocus2026, iagent2025, decoding2024, mitigatingpropensity2025, refining2026} \\

& & Amazon Movies and TV
& \citep{bifair2025, facter2025, iagent2025} \\

& & Amazon Video Games
& \citep{bifair2025, dualdebiasing2025, llmrecg2025, leadfairrec2025, towardfair2026, decoding2024} \\

& & Amazon CDs and Vinyl
& \citep{sprec2025, dualdebiasing2025, decoding2024} \\

& & Amazon Toys and Games
& \citep{dualdebiasing2025, llmfocus2026, decoding2024} \\

& & Amazon Musical Instruments
& \citep{llmrecg2025, decoding2024} \\

& & Amazon Clothing, Shoes and Jewelry
& \citep{llmfocus2026} \\

& & Amazon Industrial and Scientific
& \citep{llmrecg2025} \\

& & Amazon Sports and Outdoors
& \citep{decoding2024} \\

\cmidrule(lr){2-4}

& \multirow{3}{=}{Service and Community Platforms}
& Yelp
& \citep{towardunderstandbias2023, iagent2025, mitigatingpropensity2025} \\

& & BeerAdvocate
& \citep{leadfairrec2025} \\

& & AliEC / ZhihuRec
& \citep{comparative2025} \\

\cmidrule(lr){2-4}

& \multirow{3}{=}{High-Stakes Decision Domains}
& MIND
& \citep{gptnews2023, studyunfairness2024, fairnessbeprompted2026} \\

& & CareerBuilder
& \citep{fairwork2025, studyunfairness2024, fairnessbeprompted2026} \\

& & Insurance
& \citep{up52024, lightweight2026} \\

\midrule

\multirow{4}{=}{\textbf{Constructed Evaluation Scenario Datasets}}
& \multirow{3}{=}{Simulated Environments}
& CS-Domain Job Recommendation Data
& \citep{deconflating2026} \\

& & SynthTRIPS
& \citep{collabrec2026} \\

& & University-Profile Data
& \citep{academicbias2025} \\

\cmidrule(lr){2-4}

& Prompt-Synthesized Data
& Prompt-Synthesized Books / Movies / Songs Recommendation Data
& \citep{unveilbias2026} \\

\bottomrule
\end{tabularx}
\caption{Dataset-level mapping of fairness studies in LLM4Rec. Datasets are grouped according to the taxonomy in Figure~\ref{fig:dataset}, with representative studies listed for each dataset or data source.}

\label{tab:appendix_dataset_summary}
\end{table*}

\section{Literature Search Methodology}
\label{app:search_criteria}

\paragraph{Literature Search.}
To ensure focused and reproducible coverage of the literature, we followed a structured paper collection process. Our primary search platforms were Google Scholar, arXiv, and OpenReview. AI-assisted search was also used as a supplementary discovery tool to identify potentially missed papers; all such candidates were subsequently checked manually before inclusion.

The initial search was conducted in March 2026, followed by continuous updates as new relevant work appeared, with the final search conducted in April 2026. We used combinations of keywords related to LLM-based recommendation and fairness, including terms such as \textit{``large language model,'' ``LLM,'' ``recommender system,'' ``recommendation,'' ``fairness,'' ``bias,''} and \textit{``discrimination.''} We also used combinations targeting specific aspects covered by the survey, such as fairness evaluation, mitigation, explainability, privacy, robustness, and controllability in LLM-based recommender systems.

We considered studies from major conferences in recommender systems, natural language processing, information retrieval, data mining, and machine learning, including RecSys, ACL, EMNLP, EACL, COLING, SIGIR, WWW, WSDM, CIKM, ICML, and ICLR, as well as relevant journals in these areas. Given the rapid development of LLM4Rec and fairness research, we also included relevant preprints and under-review papers from arXiv and OpenReview when they directly addressed fairness-related issues in LLM4Rec. Our main collection focuses on papers published or released between 2022 and 2026, when fairness-oriented studies on LLM4Rec began to emerge.     Earlier works are included when they provide necessary background or are relevant to positioning our survey relative to adjacent research areas.

\paragraph{Screening and Inclusion Criteria.}
The search process yielded 101 candidate papers. Duplicate records were consolidated before screening. We then screened the candidate studies based on their titles, abstracts, and, when necessary, full texts to determine whether they directly contributed to the scope of this survey.

We included a paper if it satisfied at least one of the following conditions: (1) it explicitly studies fairness, bias, or discrimination in LLM4Rec; (2) it proposes evaluation protocols, datasets, or metrics that can be used to assess fairness-related behavior in LLM4Rec; (3) it introduces mitigation methods for reducing unfairness, popularity bias, exposure disparity, or demographic bias in LLM-enhanced recommendation pipelines; or (4) it discusses trustworthy LLM4Rec issues that are closely related to fairness, such as explainability, privacy, robustness, or controllability.

We excluded papers that focus only on traditional recommender-system fairness without involving LLMs, unless they provide necessary background for fairness definitions, evaluation protocols, or mitigation strategies. We also excluded general LLM fairness studies that do not involve recommendation scenarios, as well as LLM4Rec papers that only report recommendation accuracy without analyzing fairness-related outcomes. For papers that use highly similar datasets or metrics, we retained representative studies that best illustrate the corresponding fairness mechanism, evaluation protocol, or mitigation strategy. After screening, 77 papers were retained and incorporated into the survey taxonomy and analysis.

\section{Positioning Relative to Prior Surveys}
\label{sec:appendix_prior_surveys}

To clarify the scope and positioning of our survey, we compare it with 15 representative surveys from four adjacent research areas: fairness in recommender systems, fairness in general-purpose LLMs, trustworthy recommender systems, and LLM4Rec. As summarized in Table~\ref{tab:survey_comparison}, prior surveys typically treat only a subset of LLMs, recommendation, fairness, and cross-cutting trustworthiness as their primary scope. In contrast, our survey centers on fairness in LLM4Rec while systematically examining its connections with explainability, privacy, robustness, and controllability.

\section{Dataset-Level Mapping of Fairness Evaluation Resources}
\label{sec:appendix_dataset}

This appendix provides a dataset-level mapping of the experimental resources used in existing studies on fairness in LLM4Rec. While Section~\ref{sec:datasets} introduces the main dataset categories, Table~\ref{tab:appendix_dataset_summary} lists the individual datasets and data sources used by each line of work. The table follows the same taxonomy as Figure~\ref{fig:dataset}: curated candidate-pool datasets, behavioral interaction datasets, and constructed evaluation scenario datasets. It is intended to help readers identify which datasets have been adopted in prior studies and how they are distributed across different dataset types.

\section{Fairness Evaluation Protocols}
\label{sec:appendix_protocols}

This appendix provides a detailed view of fairness evaluation protocols in LLM4Rec. We first summarize representative metric families and studies for each protocol, followed by a comparative analysis of their assumptions, strengths, limitations, and suitable application settings.

\subsection{Protocol-Level Mapping of Fairness Evaluation Metrics}
\label{sec:appendix_protocol_metrics}

\begin{table*}[t]
\centering
\small

\begin{tabularx}{\textwidth}{p{0.08\textwidth} p{0.24\textwidth} p{0.36\textwidth} X}
\toprule
\textbf{Protocol} & \textbf{Metric Family} & \textbf{Representative Metrics} & \textbf{Representative Studies} \\
\midrule

\multirow{5}{=}{\textbf{Sensitive Attribute}}
& List similarity and ranking divergence
& Jaccard@K / IoU@K, SERP$^{*}$@K / SERP@K, PRAG$^{*}$@K / PRAG@K, Kendall's $\tau$
& \citep{gptfair2023, faireval2025, uncertaintyfairness2026, revealbiases2025, fairnessmatter2024} \\

& Sensitive-to-neutral similarity aggregation
& SNSR@K, SNSV@K, PAFS@K
& \citep{gptfair2023, faireval2025, uncertaintyfairness2026} \\

& Attribute-conditioned distribution shift
& Item-attribute distribution, category/price distribution, attribute--item association, JSD
& \citep{towardunderstandbias2023, unveilbias2026} \\

& Group-level recommendation fairness
& DP@K, EO@K, SPD, DI, EOD
& \citep{fairnessidentification2025} \\

& Representation-level attribute leakage
& Sensitive-attribute prediction accuracy / AUC
& \citep{lightweight2026} \\

\midrule

\multirow{4}{=}{\textbf{Preference Aligned}}
& Preference-aware ranker deviation
& Jaccard@K, PRAG@K, NSD, NCSD, IF, benefit deviation
& \citep{cfairllm2025, normative2024} \\

& Group benefit and fairness disparity
& $\Delta B$, benefit deviation, DP@K, EO@K
& \citep{cfairllm2025, normative2024, fairaugment2025, improvefairness2025, comparative2025} \\

& Utility--fairness trade-off metrics
& Precision@K, Recall@K, HR@K, NDCG@K
& \citep{fairaugment2025, improvefairness2025, investigatingmitigatingunfairness2025, leadfairrec2025} \\

& Agentic preference and user-protection alignment
& FR@K, P-HR@K, P-NDCG@K, P-MRR@K, IoI, IoR, Accuracy, Acceptance, Coherence
& \citep{iagent2025, refining2026} \\

\midrule

\multirow{4}{=}{\textbf{Target Specific}}
& Group parity and counterfactual ranking disparity
& SP, EO, PPV\_diff, Pred\_diff, U-Metric, U-NDCG@K
& \citep{fairwork2025, studyunfairness2024} \\

& Sensitive-to-neutral output similarity
& Jaccard@K, SERP@K, PRAG@K, BERTScore, SNSR@K, SNSV@K
& \citep{fairnessbeprompted2026} \\

& Domain representation and content-bias metrics
& DRS, GRS, polarization score, toxicity score, sentiment score, political-leaning and demographic exposure
& \citep{academicbias2025, polarization2026} \\

& Agentic bias robustness and constraint-aware ranking
& Acc$_{ori}$, Acc$_{inj}$, BSR, RR, Recall@K, NDCG@K under preference and qualification tasks
& \citep{llmasarecommende2026, deconflating2026} \\

\midrule

\multirow{4}{=}{\textbf{Item Side}}
& Exposure concentration and provider fairness
& Gini Index, HHI, Entropy, catalog coverage, popular-provider Precision@K, self-promotion rate
& \citep{understandingbiases2025, gptnews2023, dubash2026self} \\

& Popularity bias and long-tail exposure
& ARP@K, APT@K, Long-tail Coverage, PL, SPL, recommendation frequency by popularity group
& \citep{towardfair2026, dualdebiasing2025, split2026, semanticandbias2026} \\

& Group-level item-side unfairness
& GU@K, MGU@K, DGU@K, item-group fairness gap, category-level unfairness
& \citep{item-sidefair2024, sprec2025, bifair2025} \\

& Diversity, homogeneity, and system-induced bias
& DivRatio, ORRatio, BLEU, category entropy, category repetition ratio, exposure polarization, moderator success rate
& \citep{sprec2025, decoding2024, llmfocus2026, echoes2026, collabrec2026} \\

\bottomrule
\end{tabularx}
\caption{Protocol-level mapping of fairness metrics in LLM4Rec. Metrics are grouped according to the evaluation protocol taxonomy in Figure~\ref{fig:benchmark}, with representative studies listed for each metric family.}
\label{tab:appendix_protocol_metrics}
\end{table*}

Table~\ref{tab:appendix_protocol_metrics} presents a protocol-level overview of representative fairness metrics used in LLM4Rec evaluation. Complementing the discussion in Section~\ref{sec:evaluation}, it organizes existing metrics into representative families under four evaluation protocols: sensitive-attribute, preference-aligned, target-specific, and item-side protocols. Rather than serving as an exhaustive inventory of all metric variants, the table aims to clarify the main evaluation focus of each protocol and to identify representative studies that exemplify each metric family.

\paragraph{Sensitive Attribute}
Sensitive-attribute evaluation protocols examine whether recommendation outputs change when demographic, social, personality-related, or other sensitive attributes are introduced into otherwise comparable prompts. As shown in Table~\ref{tab:appendix_protocol_metrics}, these metrics mainly capture list-level changes, aggregated sensitive-to-neutral stability, shifts in recommended item distributions, group-level outcome disparities, and representation-level attribute leakage. Together, they evaluate whether sensitive attributes influence either the visible recommendation results or the latent representations used by the system.

\paragraph{Preference Aligned}
Preference-aligned evaluation protocols assess fairness with respect to user preference satisfaction and recommendation benefit. Instead of treating every output difference as unfair, they ask whether recommendation changes lead to unjustified deviations from preference-consistent rankings or unequal utility across user groups. The metric families in Table~\ref{tab:appendix_protocol_metrics} therefore include ranker deviation, group benefit disparity, utility--fairness trade-off metrics, and agentic alignment measures for interactive recommendation settings. Standard utility metrics such as Precision@K and Recall@K are often reported alongside fairness metrics to assess whether disparity reduction preserves recommendation quality.

\paragraph{Target Specific}
Target-specific evaluation protocols are used in domains where fairness assumptions depend strongly on the application context, such as job recommendation, academic recommendation, political content curation, and agentic decision-making. Their metrics measure whether sensitive attributes or contextual changes lead to unequal decisions, ranking disparities, biased representation, or vulnerability to injected biases. As summarized in Table~\ref{tab:appendix_protocol_metrics}, these protocols often combine general fairness metrics with domain-specific measures of representation, content bias, robustness, and constraint satisfaction.

\paragraph{Item Side}
Item-side evaluation protocols focus on whether recommendation exposure is fairly distributed across items, providers, popularity groups, categories, or other item-side stakeholders. The metric families in Table~\ref{tab:appendix_protocol_metrics} cover exposure concentration, provider fairness, popularity bias, long-tail exposure, item-group unfairness, diversity, homogeneity, and system-induced bias. These metrics are especially important for diagnosing whether LLM4Rec over-expose popular or dominant items while suppressing long-tail or underrepresented alternatives.

\subsection{Comparative Analysis of Fairness Evaluation Protocols}
\label{sec:appendix_protocol_comparison}

While Table~\ref{tab:appendix_protocol_metrics} summarizes the metric families and representative studies associated with each evaluation protocol, the protocols also differ substantially in their underlying assumptions, interpretation of fairness, and appropriate application settings. Table~\ref{tab:appendix_protocol_comparison} therefore provides a complementary comparison along these dimensions, highlighting the practical strengths and limitations of each protocol.

\begin{table*}[t]
\centering
\small
\setlength{\tabcolsep}{3pt}
\renewcommand{\arraystretch}{1.15}

\begin{tabularx}{\textwidth}{
p{0.1\textwidth}
p{0.12\textwidth}
p{0.145\textwidth}
p{0.165\textwidth}
p{0.22\textwidth}
X}
\toprule
\textbf{Protocol}
& \textbf{Core Question}
& \textbf{Key Assumption}
& \textbf{Strengths}
& \textbf{Limitations}
& \textbf{Best Suited For} \\
\midrule

\textbf{Sensitive Attribute}
&
Does changing only a sensitive attribute in the user profile or prompt alter the recommendation output?
&
Under the controlled counterfactual, the modified attribute is assumed to be normatively irrelevant once all preference-relevant information is held constant.
&
Simple and controlled counterfactual audit; requires little or no interaction ground truth; directly captures prompt sensitivity and differential treatment.
&
Output differences do not necessarily indicate unfairness, as they may reflect legitimate preference differences or stochastic generation. Results can also be sensitive to prompt templates, decoding settings, and similarity metrics.
&
Controlled audits of prompt-based or zero-shot LLM4Rec, especially when preference labels or utility ground truth are unavailable.
\\

\midrule

\textbf{Preference Aligned}
&
Do users or groups receive comparable recommendation benefit after accounting for their actual preferences?
&
User preferences and recommendation benefit can be estimated reliably and compared meaningfully across groups.
&
Distinguishes harmful disparity from benign output variation; more directly reflects user welfare than list-level invariance; avoids treating every demographic difference as unfair.
&
Requires reliable preference or benefit estimation, which may be unavailable or model-dependent. It may overlook provider-side harms and disparities not captured by the selected utility function.
&
Consumer-side fairness evaluation when interaction histories, preference labels, relevance judgments, or defensible benefit functions are available.
\\

\midrule

\textbf{Target Specific}
&
Does an LLM4Rec produce equitable decisions or rankings in a particular high-stakes or domain-specific setting?
&
The domain provides meaningful qualification, relevance, or outcome criteria against which fairness can be assessed.
&
Supports realistic and decision-relevant audits; can capture proxy discrimination and domain-specific harms; accommodates counterfactual and group-parity analysis.
&
Findings and metrics may not generalize across domains. Fairness definitions can depend on contested normative choices, and observed differences may be confounded by legitimate qualification differences.
&
Domain-specific recommendation or matching tasks, such as employment, education, insurance, and other settings with explicit decision criteria.
\\

\midrule

\textbf{Item Side}
&
Are inclusion, exposure, and recommendation opportunities distributed equitably across items, providers, or popularity groups?
&
Exposure is a meaningful proxy for item or provider benefit, and item groups such as head and long-tail items can be defined appropriately.
&
Directly evaluates provider fairness, popularity amplification, concentration, and long-tail suppression; complements user-side utility evaluation.
&
Greater exposure equality does not necessarily imply fair or relevant recommendations. Results depend on catalog construction, group definitions, ranking-position weights, and candidate availability, and improvements may trade off with user utility.
&
Ranking, re-ranking, and generative recommendation settings where provider opportunity, catalog diversity, popularity bias, or long-tail exposure is central.
\\

\bottomrule
\end{tabularx}

\caption{Comparison of fairness evaluation protocols in LLM4Rec. The protocols differ not only in their evaluation targets, but also in their underlying assumptions, interpretability, limitations, and suitable application settings.}
\label{tab:appendix_protocol_comparison}
\end{table*}

\section{Comparative Analysis of Fairness Mitigation Strategies}
\label{sec:appendix_mitigation_comparison}

Section~\ref{sec:mitigation} organizes fairness mitigation strategies according to where interventions are applied in the LLM4Rec pipeline. While this pipeline-level taxonomy clarifies the mechanism of each strategy, the intervention levels also differ in their assumptions, deployment requirements, persistence, and potential trade-offs. Table~\ref{tab:appendix_mitigation_comparison} therefore provides a comparative view of the four intervention levels, highlighting their practical strengths, limitations, and suitable deployment settings.

\begin{table*}[t]
\centering
\small
\setlength{\tabcolsep}{3pt}
\renewcommand{\arraystretch}{1.15}

\begin{tabularx}{\textwidth}{
p{0.07\textwidth}
p{0.125\textwidth}
p{0.17\textwidth}
p{0.18\textwidth}
p{0.215\textwidth}
X}
\toprule
\textbf{Stage}
& \textbf{Intervention Mechanism}
& \textbf{Key Assumption}
& \textbf{Strengths}
& \textbf{Limitations}
& \textbf{Best Suited For} \\
\midrule

\textbf{Input Level}
&
Prompt engineering, prompt optimization, inference-time guidance, and conformal thresholding
&
Fairness can be improved by steering the LLM at inference time without modifying model parameters.
&
Requires no retraining; low deployment cost; easy to integrate; compatible with closed-source and API-based LLMs.
&
Sensitive to prompt wording and model capability; improvements may be unstable or fail to generalize across tasks and models; cannot directly remove deeper biases encoded in data or model parameters.
&
API-based or closed-source LLMs, rapid deployment, and settings where model retraining or parameter access is infeasible.
\\

\midrule

\textbf{Data Level}
&
Counterfactual data augmentation, interaction-history perturbation, and causal intervention
&
A substantial source of unfairness arises from biased interaction data, exposure patterns, selection effects, or popularity distributions.
&
Addresses bias upstream before recommendation generation; can benefit downstream models; may improve both user-side utility and item-side exposure fairness.
&
Requires access to training or interaction data; effectiveness depends on counterfactual quality and causal assumptions; may not address biases introduced by prompting, model representations, or decoding.
&
Systems with accessible interaction logs and fairness problems primarily driven by popularity, exposure, propensity, or selection bias.
\\

\midrule

\textbf{Model Level}
&
PEFT, adapters, kernelized projection, fairness regularization, bi-level optimization, and mixture-of-experts
&
Fairness can be improved by modifying internal representations, model parameters, architectures, or optimization objectives.
&
Provides more persistent mitigation than inference-time prompting; directly encodes fairness objectives or constraints; can support multi-attribute and representation-level debiasing.
&
Requires model access and additional training; may introduce fairness--utility trade-offs; effectiveness may not transfer reliably across tasks, datasets, or domains.
&
Open-source or internally controlled models where stronger and more persistent fairness mitigation is required.
\\

\midrule

\textbf{Output Level}
&
Fairness-aware decoding, post-hoc re-ranking, explainable re-rankers, and IPS-based debiasing
&
Bias is introduced or amplified during candidate ordering, decoding, or final output generation and can therefore be corrected at the output stage.
&
Directly controls final recommendation outputs; can be integrated without retraining the base model; particularly effective for exposure, popularity, and provider-side fairness.
&
Limited by the quality and diversity of the candidate pool; may trade off recommendation relevance against fairness; often mitigates downstream manifestations rather than upstream sources of bias.
&
Item- or provider-side fairness, long-tail exposure, and systems with fixed retrieval components or base models where output-stage intervention is feasible.
\\

\bottomrule
\end{tabularx}

\caption{Comparison of fairness mitigation strategies across the LLM4Rec pipeline. Intervention levels differ in where fairness is imposed, what system access they require, how persistent their effects are, and which fairness problems they are best suited to address.}
\label{tab:appendix_mitigation_comparison}
\end{table*}

\end{document}